\documentclass[a4paper,11pt]{article}
\usepackage{pos}
\usepackage{amsmath}

\graphicspath{{./plots/}}

\newcommand{\lr}[1]{ \left( #1 \right) }
\newcommand{\lrs}[1]{ \left[ #1 \right] }
\newcommand{\lrc}[1]{ \left\{ #1 \right\} }
\newcommand{\dt}[0]{{\Delta \tau}}

\newcommand{\tr}[0]{\mathrm{tr}}

\newcommand{\vev}[1]{ \left\langle \, #1 \, \right\rangle }

\newcommand{\ket}[1]{ \, | #1 \rangle }
\newcommand{\bra}[1]{ \langle #1 | \, }

\newcommand{\abs}[1]{| #1 |}

\title{Hamiltonian-based dimensional reduction and spectral reconstruction with Wilson-Dirac fermions}
\ShortTitle{Hamiltonian dimensional reduction with Wilson-Dirac fermions}

\author*[a]{P. Buividovich}
\author[a]{B. Hind}

\affiliation[a]{Department of Mathematical Sciences, University of Liverpool, \\
L69 7ZX Liverpool, UK}

\emailAdd{ben.hind@liverpool.ac.uk}
\emailAdd{pavel.buividovich@liverpool.ac.uk}

\abstract{Motivated by the process of reconstructing real-time spectral functions from Euclidean correlators in lattice QCD, we derive explicit expressions for the fermionic determinant and the propagator of the four-dimensional clover-improved Wilson-Dirac fermions on anisotropic lattices in terms of the three-dimensional Wilson-Dirac Hamiltonian operator. We derive an effective Hamiltonian that governs Euclidean time evolution at finite temporal lattice spacing, and demonstrate its hermiticity and particle-anti-particle symmetry. Our results allow to quantify lattice artifacts of the numerical spectral reconstruction based on Euclidean fermionic correlators at finite temporal lattice spacing.}

\FullConference{The 42nd International Symposium on Lattice Field Theory (LATTICE2025)\\
2-8 November 2025\\
Tata Institute of Fundamental Research, Mumbai, India\\}

\begin{document}
\sloppy
\maketitle

\section{Introduction}
\label{sec:intro}

The fermionic determinant and fermionic Green's functions in four-dimensional Euclidean space play a central role in numerical simulations of strongly interacting fermions. Via numerical spectral reconstruction based on the Green-Kubo relations, correlators of fermionic bilinear operators (such as electric current or axial charge) provide first-principle insights into charge transport properties in strongly interacting quantum field theories and many-body systems. 

In lattice QCD, these objects are usually calculated in terms of some discretization of the four-dimensional Euclidean Dirac operator $D = \gamma_{\mu} \lr{\partial/\partial x_{\mu} - i A_{\mu}} + m$, where $\gamma_{\mu} = \gamma_{\mu}^{\dagger}$ are Euclidean gamma-matrices, $A_{\mu} = \lrc{A_0, \vec{A}}$ are gauge fields, and $m$ is the quark mass. In some applications, such as simulations of canonical partition functions at finite fermion density \cite{Nakamura:1009.2149,Alexandru:1009.2197}, spectral reconstruction problems \cite{Buividovich:25:3}, or numerical studies of anomalous transport phenomena within the Wigner-Weyl formalism \cite{zubkov2018momentum,chernodub2017scale,zhang2020influence}, it is advantageous to work with dimensionally reduced operators in three spatial dimensions. In particular, a three-dimensional representation of fermionic observables and path integral weight can be related to the Hamiltonian formulation of gauge theories \cite{Kogut:75:1}, where many questions related to real-time dynamics, transport phenomena, quantum entanglement, and finite fermion density can be asked in a more natural way. The Hamiltonian formulation is also attracting considerable attention in the context of simulating gauge theories on quantum computers.

The fermionic part of the Hamiltonian of a lattice gauge theory has a general form $\hat{H}_F = \sum\limits_{\vec{x}, \vec{y}, \alpha, \beta} \hat{\psi}^{\dag}_{\vec{x},\alpha} h_{\vec{x}, \alpha; \vec{y}, \beta}\lrs{\vec{A}} \hat{\psi}_{\vec{y}, \beta}$, where summation goes over spatial lattice sites $\vec{x}$ and $\vec{y}$ and any internal indices $\alpha, \beta$ such as spin, colour, or flavour. $\hat{\psi}^{\dag}_{\vec{x}, \alpha}$ and $\hat{\psi}_{\vec{y}, \beta}$ are the fermionic creation/annihilation operators, and $h_{\vec{x}, \alpha; \vec{y}, \beta}\lrs{\vec{A}}$ is a single-particle fermionic Hamiltonian that depends on spatial components of the gauge field $\vec{A}$. It is convenient to use the short-hand index-free bilinear form notation $\hat{H}_F = \hat{\psi}^{\dag} h \psi$.

The partition function of lattice gauge theory can be directly represented as a path integral over Euclidean-space gauge fields $\lrc{A_0\lr{\tau, \vec{x}}, \vec{A}\lr{\tau, \vec{x}}}$ with the weight that includes the fermionic determinant with the single-particle Hamiltonian $h$ \cite{Blankenbecler:PhysRevD.24.2278,Buividovich:12:1}:
\begin{eqnarray}
\label{eq:3d_determinant_generic}
 \mathcal{Z} 
 = 
 \int \mathcal{D} A_0\lr{\tau, \vec{x}} \, \mathcal{D} \vec{A}\lr{\tau, \vec{x}} e^{-S_{YM}\lrs{A_0, \vec{A}}}
 \det\lr{
    1 + \prod\limits_{\tau = 0}^{N_{\tau} - 1} T_{\tau} } ,
\end{eqnarray}
where $S_{YM}\lrs{A_0, \vec{A}}$ is the bosonic action for gauge fields, $T_{\tau} = e^{-\dt \, h\lrs{\vec{A}\lr{\tau}}} \, e^{i \dt \, A_0\lr{\tau}}$ are the fermionic transfer matrices, and $e^{-\dt \, h\lrs{\vec{A}\lr{\tau}}}$ are matrix exponentials of the single-particle Hamiltonian $h$ with the gauge field $\vec{A}\lr{\tau} \equiv \vec{A}\lr{\tau, \vec{x}}$ that depends on Euclidean time $\tau$. The latter is discretized into $N_{\tau}$ time slices separated by small steps $\dt$. The factors $e^{i \dt \, A_0\lr{\tau} }$ in (\ref{eq:3d_determinant_generic}) are holonomy operators that parallel transport fermionic wave functions between neighbouring time slices and include only the time-like components $A_0\lr{\tau, x}$ of gauge fields. 

Connected Euclidean correlators of fermionic bilinear operators 
$\hat{\mathcal{O}} = \sum\limits_{\vec{x}, \vec{y}, \alpha, \beta} \hat{\psi}^{\dag}_{\vec{x},\alpha} O_{\vec{x}, \alpha; \vec{y}, \beta} \hat{\psi}_{\vec{y}, \beta}$ can be likewise represented in terms of $\tau$-ordered products of $T_{\tau}$:
\begin{eqnarray}
\label{eq:bilinear_correlator_general}
 G\lr{\tau} = \mathcal{Z}^{-1} 
 \tr\lr{\hat{\mathcal{O}} e^{-\tau \hat{H}} \hat{\mathcal{O}} e^{-\lr{\beta - \tau} \hat{H}} }
 = \nonumber \\ = 
 \vev{ \tr\lr{
 O \,
 \mathcal{F}
 \prod\limits_{\tau'=0}^{\tau - \dt} T_{\tau'}
 \, O  \,
 \prod\limits_{\tau'=\tau}^{N_{\tau}-1} T_{\tau'}
 \mathcal{F}
 }} ,
\end{eqnarray}
where $\hat{H}$ is the full quantum Hamiltonian, $\mathcal{Z} = \tr \lr{e^{-\beta \hat{H}}}$ is the partition function, $\mathcal{F} = \lr{1 + \prod\limits_{\tau' = 0}^{N_{\tau}-1} T_{\tau'}}^{-1}$ is a generalization of the Fermi factor $\lr{1 + e^{-\beta h}}^{-1}$, where $\beta \equiv T^{-1}$ is an inverse temperature. In (\ref{eq:bilinear_correlator_general}) we average over all gauge configurations with the weight dictated by the path integral weight in the partition function (\ref{eq:3d_determinant_generic}). 

The expression (\ref{eq:bilinear_correlator_general}) becomes particularly instructive in the high-temperature, semi-classical limit, where the gauge fields become effectively $\tau$-independent, the time-like gauge field components $A_0$ decouple, and (\ref{eq:bilinear_correlator_general}) reduces to the expression that allows for direct reconstruction of real-time spectral function in terms of eigenvalues of the single-particle Hamiltonian $h \equiv h\lrs{\vec{A}}$ in the background of static gauge fields $\vec{A}\lr{\vec{x}}$ \cite{Buividovich:25:3}:
\begin{eqnarray}
\label{eq:bilinear_correlator_genera_high_temp}
 G\lr{\tau} = \tr\lr{
  O 
  \frac{1}{1 + e^{-\beta h}} e^{-\tau h} 
  O 
  e^{-\lr{\beta - \tau} h}  
  \frac{1}{1 + e^{-\beta h}} 
  } .
\end{eqnarray}

In \cite{Nakamura:1009.2149,Alexandru:1009.2197} it was demonstrated that the determinant of a four-dimensional Wilson-Dirac operator $D_{WD}$ can also be reduced to the determinant of a three-dimensional operator with a structure similar to (\ref{eq:3d_determinant_generic}): $\det\lr{D_{WD}} \sim \det\lr{1 + \prod\limits_{\tau} T_{\tau}^{WD}}$. While the properties of the fermionic transfer matrices $T_{\tau}^{WD}$ were analyzed in detail \cite{Wenger:2302.07385,Futamura:1411.4262,Nagata:2014bra,Bilgici:2009gjc,Nakamura:1009.2149,Alexandru:1009.2197}, their relation to the single-particle fermionic Hamiltonian in the general expression (\ref{eq:3d_determinant_generic}) and their Hermiticity has not been demonstrated so far. Explicit expressions for expectation values of fermionic bilinear operators (\ref{eq:bilinear_correlator_general}) are also not available in the literature.

In these Proceedings, we reformulate the dimensional reduction formulas of \cite{Nakamura:1009.2149,Alexandru:1009.2197} directly in terms of the Wilson-Dirac Hamiltonian, and extend them to the case of an anisotropic fermionic action with different lattice spacing for temporal and spatial directions \cite{Chen:hep-lat/0006019,Aarts:1412.6411}. This allows us to directly expose time discretization artifacts in Euclidean-time fermionic correlators. These artifacts directly affect numerical estimates of spectral functions obtained by inverting Green-Kubo relations on Euclidean correlators in lattice QCD.

\section{Dimensional reduction for the anisotropic clover-improved Wilson-Dirac operator}
\label{sec:dimensional_reduction}

Our starting point is the clover-improved Wilson-Dirac operator $D_{\text{WD}}$ \cite{Chen:hep-lat/0006019,Aarts:1412.6411}, implemented in the \texttt{OpenQCD-FASTSUM} code \cite{OpenQCDfastsum}:
\begin{eqnarray}
\label{eq:DWD}
    \lr{D_{WD}}_{\tau, \vec{x}; \tau', \vec{y}} 
    = 
    \frac{1}{u_t} \, \lr{ m_0 \, u_t + 1 + \frac{3}{\gamma_f} + C^{t}_{\tau, \vec{x}} + \frac{1}{\gamma_f} C^{s}_{\tau, \vec{x}} } 
    \delta_{\vec{x},\vec{y}} \delta_{\tau, \tau'} 
    - \nonumber \\ -
    \frac{1}{u_t} \lr{ P_{-} \, U_{\tau, \vec{x}, 0} \, s_{\tau} \, \delta_{\tau + 1, \tau'}  +  P_{+} \, U^{\dagger}_{\tau', \vec{y}, 0} \, s_{\tau'} \, \delta_{\tau -1, \tau'} } \delta_{\vec{x}, \vec{y}} 
    - \\ \nonumber -
    \frac{1}{\gamma_f \, u_t}  \sum_{k=1}^{3} \lr{ \lr{ \frac{1 - \gamma_k}{2} } U_{\tau, \vec{x}, k} \delta_{\vec{x}+\vec{e}_k, \vec{y}}  +  \lr{ \frac{1 + \gamma_k}{2} } U^{\dagger}_{\tau', \vec{y}, k} \delta_{\vec{x}-\vec{e}_k, \vec{y}} } \delta_{\tau, \tau'} ,    
\end{eqnarray}
where $m_0$ is the quark mass, $u_t$ is the tadpole improvement factor, and $U_{\tau, \vec{x}, \mu}$ are $SU\lr{3}$-valued link variables. The factor $s_{\tau}$ is equal to $+1$ for all $\tau = 0 \ldots N_{\tau} - 2$, and is set to $-1$ for the last time slice with $\tau = N_{\tau} - 1$ to account for anti-periodic boundary conditions for fermions. $\gamma_f$ is the bare fermionic anisotropy which can be interpreted as an unrenormalized ratio $a_s/\dt$ of spatial $a_s$ and temporal $\dt$ lattice spacings, hence the continuum-time limit corresponds to the limit $\gamma_f \rightarrow +\infty$. $\vec{e}_{k}$ are the basis vectors on the spatial lattice that correspond to shifts by one lattice spacing in the direction $k$. We assume periodic boundary conditions for spatial and temporal directions. In particular, index shifts in $\delta_{\tau\pm1, \tau'}$ assume that $\tau$ is a cyclic variable defined modulo the temporal lattice size. $P_{\pm} = \frac{1 \pm \gamma_0}{2}$ are the spinor projection operators with the following properties:
\begin{eqnarray}
\label{eq:projector_properties}
    \gamma_0 P_{\pm} = \pm P_{\pm}, 
    \quad 
    P_{\pm}^2 &= P_{\pm}, 
    \quad
    P_{\pm} P_{\mp} = 0, 
    \quad
    P_{\pm} + P_{\mp} = 1 .
\end{eqnarray}
$C^{t}_{\tau, \vec{x}}$ and $C^{s}_{\tau, \vec{x}}$ are the temporal and spatial clover terms:
\begin{eqnarray}
\label{eq:clover_terms}
    C^{t}_{\tau, \vec{x}}  = -\frac{1}{2} \frac{c_T}{u_t u_s^2} \sum_{k=1}^{3} \frac{i}{2} \lrs{\gamma_0, \gamma_k} F_{\tau, \vec{x}, 0k}, 
    \\
    C^{s}_{\tau, \vec{x}} = -\frac{1}{2} \frac{c_R}{\nu u_s^3} \sum_{1\le j < k \le 3} \frac{i}{2} \lrs{\gamma_j, \gamma_k} F_{\tau, \vec{x}, j k} ,
\end{eqnarray}
where $c_T$ and $c_R$ are the anisotropic temporal and spatial clover parameters, $\nu$ is the ratio of bare gluonic and fermionic anisotropies, $u_s$ is the spatial tadpole improvement factor, and $F_{\vec{x}, \tau, \mu \nu} = F^{\dag}_{\vec{x}, \tau, \mu \nu} = -F_{\vec{x}, \tau, \nu \mu}$ is the lattice field strength tensor that takes values in $SU\lr{3}$ Lie algebra (Hermitian, traceless matrices).

We now introduce the Wilson-Dirac single-particle Hamiltonian
\begin{eqnarray}
\label{eq:hWD}
    \lr{h^{WD}_{\tau}}_{\vec{x}, \vec{y}} = \gamma_0 \lr{ m_0 \, u_t \, \gamma_f + 3 + C^{s}_{\tau, \vec{x}} } \delta_{\vec{x},\vec{y}}  
    - \\ \nonumber - \gamma_0 \, \sum_{k=1}^{3} \lr{ \lr{ \frac{1 - \gamma_k}{2} } U_{\tau, \vec{x}, k} \delta_{\vec{x}+\vec{e}_k, \vec{y}}  +  \lr{ \frac{1 + \gamma_k}{2} } U^{\dagger}_{\tau', \vec{y}, k} \delta_{\vec{x}-\vec{e}_k, \vec{y}} } 
\end{eqnarray}
that depends on the Euclidean time $\tau$ via the link variables $U_{\tau, \vec{x}, k}$. $h^{WD}$ is a Hermitian operator. In continuous spacetime, the four-dimensional Dirac operator can be related to the $\tau$-dependent single-particle Dirac Hamiltonian $h_{\tau}$ as $D = \gamma_0 \lr{\partial_{\tau} + h_{\tau}}$, or, equivalently, $\gamma_0 \, D = \partial_{\tau} + h_{\tau}$. Motivated by this relation, we rewrite the product $u_t \, \gamma_0 \, D_{WD}$ as
\begin{eqnarray} 
\label{eq:g0DWD}
    u_t \, \lr{\gamma_0 D_{WD}}_{\tau, \vec{x}; \tau', \vec{y}} 
    = 
    \gamma_0 \lr{1 + C^{t}_{\tau, \vec{x}} } \delta_{x,y} \delta_{\tau, \tau'}
    + \nonumber \\ +
    P_{-} U_{\tau, \vec{x}, 0}  \delta_{\vec{x}, \vec{y}} \, s_{\tau} \, \delta_{\tau+1, \tau'}
    -  
    P_{+} U^{\dagger}_{\tau', \vec{y}, 0} \delta_{\vec{x}, \vec{y}} \, s_{\tau'} \, \delta_{\tau-1, \tau'}
    +
    \gamma_f^{-1} \, \lr{h^{WD}_{\tau}}_{\vec{x}, \vec{y}} \, \delta_{\tau, \tau'} .
\end{eqnarray}
Following \cite{Alexandru:1009.2197,Nakamura:1009.2149}, we now define an auxiliary four-dimensional operator
\begin{eqnarray}
\label{eq:P_aux}
    \lr{\mathcal{P}}_{\tau, \vec{x}; \tau', \vec{y}}  
    = 
    P_{-}  \delta_{\tau, \tau'} \delta_{\vec{x}, \vec{y}} + P_{+} U_{\tau, \vec{x}, 0} \, s_{\tau} \, \delta_{\tau + 1, \tau'} \delta_{\vec{x}, \vec{y}}
\end{eqnarray}
with $\det \mathcal{P} = 1$.

The product of $u_t \, \gamma_0 \, D_{WD}$ and $\mathcal{P}$ can be represented as a block matrix with blocks indexed by Euclidean time labels $\tau$, $\tau'$, where the blocks are only non-zero on the main diagonal with $\tau' = \tau$ and for $\tau' = \tau+1$ (up to periodic boundary conditions):
\begin{eqnarray}
\label{eq:g0_Dwd_P}
 u_t \, \lr{\gamma_0 D_{WD} \mathcal{P}}_{\tau, \vec{x}; \tau', \vec{y}} 
 = 
 - \lr{\alpha_{\tau}}_{\vec{x}, \vec{y}} \, \delta_{\tau, \tau'}
 + 
 \lr{\beta_{\tau}}_{\vec{x}, \vec{y}} \, U_{\tau, \vec{y}, 0} \, s_{\tau} \, \delta_{\tau+1, \tau'}
 \quad ,
\end{eqnarray}
where we introduced the three-dimensional operators 
\begin{eqnarray}
\label{eq:alpha_beta}
 \lr{\alpha_{\tau}}_{\vec{x}, \vec{y}}
 = 
 \delta_{\vec{x}, \vec{y}} 
  - C^{t}_{\tau, \vec{x}} \, P_{-} \, \delta_{\vec{x}, \vec{y}}
  - \gamma_f^{-1} \, \lr{h^{WD}_{\tau}}_{\vec{x}, \vec{y}} \, P_{-} \quad ,
 \nonumber \\
 \lr{\beta_{\tau}}_{\vec{x}, \vec{y}}
 =
 \delta_{\vec{x}, \vec{y}} 
  - C^{t}_{\tau, \vec{x}} \, P_{+} \, \delta_{\vec{x}, \vec{y}}
  + \gamma_f^{-1} \, \lr{h^{WD}_{\tau}}_{\vec{x}, \vec{y}} \, P_{+} \quad .
\end{eqnarray}

The time-like clover terms were simplified using the identities $\gamma_0 C^{t}_{\tau, \vec{x}} \, P_{\pm} = - C^{t}_{\tau, \vec{x}} \, \gamma_0 \, P_{\pm} = \mp C^{t}_{\tau, \vec{x}} \, P_{\pm}$ that follow from the algebra of Euclidean gamma-matrices.

Using the identities $\det\lr{\gamma_0} = 1$ and $\det\lr{\mathcal{P}} = 1$, we can write $\det{D_{WD}} = u_t^{-N} \det\lr{u_t \, \gamma_0 \, D_{WD} \, \mathcal{P}}$, where $N$ is the overall dimension of the linear operator $D_{WD}$. Using the special ``periodic upper-diagonal'' structure of the block matrix $u_t \, \gamma_0 \, D_{WD} \, \mathcal{P}$, its determinant can now be reduced to a product of determinants of three-dimensional operators \cite{Alexandru:1009.2197,Nakamura:1009.2149}:
\begin{eqnarray}
\label{eq:det_final}
    \det\lr{D_{WD}}
    =
    u_t^{-N} \det\lr{u_t \, \gamma_0 \, D_{WD} \, \mathcal{P}}
    =
    \lr{ \prod_{\tau=1}^{N_{\tau}} \det(u_t^{-1} \, \alpha_{\tau}) }
    \,
 \det\lr{ 1 + \prod_{\tau=1}^{N_{\tau}} \lr{\alpha_{\tau}^{-1} \beta_{\tau} U_{\tau, 0} }}
 ,
\end{eqnarray}
where $N_{\tau}$ is the total number of time slices, and $U_{\tau, 0}$ is a three-dimensional operator defined as $\lr{U_{\tau, 0}}_{\vec{x}, \vec{y}} = U_{\tau, \vec{x}, 0} \, \delta_{\vec{x}, \vec{y}}$. Comparing this expression with the fermionic determinant in (\ref{eq:3d_determinant_generic}), we conclude that the three-dimensional operator $\alpha_{\tau}^{-1} \beta_{\tau} U_{\tau, 0}$ should approximate the three-dimensional operator $e^{-\dt \, h\lrs{\vec{A}\lr{\tau}}} \, e^{i \dt \, A_0\lr{\tau} }$. On the lattice, the gauge field variables $\vec{A}\lr{\tau}$ and $A_0\lr{\tau}$ are encoded in the link variables $U_{\tau, \vec{x}, k}$ and $U_{\tau, \vec{x}, 0}$. 

Let us now demonstrate that the product $\alpha_{\tau}^{-1} \, \beta_{\tau}$ indeed converges to $e^{\dt \, h^{WD}}$ in the limit $\dt \rightarrow 0$, or, equivalently, $\gamma_f \rightarrow +\infty$. By virtue of particle-anti-particle symmetry of the Dirac Hamiltonian, this transfer matrix is equivalent to $e^{-\dt \, h^{WD}}$.

It is now convenient to express all three-dimensional operators as block matrices in a ``non-relativistic'' representation for Euclidean $\gamma$-matrices, where $\gamma_0 = \mathrm{diag}\lr{1, -1}$. The Hamiltonian and the $\alpha$ and $\beta$ matrices then have the following structure (to shorten the notation, we omit the $\tau$ indices in what follows, understanding that all operators depend on the time slice $\tau$):
\begin{eqnarray}
\label{eq:hamiltonian_block}
    h^{WD} = 
    \lr{
    \begin{array}{cc}
     \mathfrak{m} &
     -i \vec{\sigma} \cdot \vec{\nabla} \\
     -i \vec{\sigma} \cdot \vec{\nabla} &
     -\mathfrak{m}
    \end{array}
    } ,
    \quad
    \alpha = 
    \lr{
     \begin{array}{cc}
     1 &
     - \gamma_f^{-1} \, \mathfrak{p}_{+} \\
     0 &
     1 + \gamma_f^{-1} \, \mathfrak{m} 
    \end{array}
    } ,
    \quad
    \beta = 
    \lr{
     \begin{array}{cc}
     1 + \gamma_f^{-1} \, \mathfrak{m}  &
     0 \\
     \gamma_f^{-1} \, \mathfrak{p}_{-} &
     1 
    \end{array}
    } ,
    \nonumber \\
    \mathfrak{m} = \mathfrak{m}^{\dag} = m_0 \, u_t \, \gamma_f - \frac{\Delta}{2} + \frac{c_R}{2 \,\nu \, u_s^3} \vec{\sigma}\cdot\vec{\mathcal{B}},
    \quad
    \mathfrak{p}_{\pm} = \mathfrak{p}_{\pm}^{\dag} = \vec{\sigma} \cdot \lr{-i \vec{\nabla} \pm \frac{c_T}{2 \, u_t \, u_s^2} \vec{\mathcal{E}}},   
\end{eqnarray}
where $\Delta$ is a covariant lattice Laplacian, $\vec{\sigma}$ is a vector of Pauli matrices, $\vec{\mathcal{B}} \equiv 1/2 \sum\limits_{i,j,k} \epsilon_{ijk} F_{ij} \vec{e}_k$ and $\vec{\mathcal{E}} \equiv \gamma_f \, F_{0k} \vec{e}_k$ are the lattice chromo-magnetic and chromo-electric fields, and $\vec{\nabla}$ is a covariant lattice derivative based on a central finite-difference approximation.
Using the general expression for the inverse of block matrices, we find
\begin{eqnarray}
\label{eq:alphainv_beta}
   \alpha^{-1} \beta = 
    \lr{
    \begin{array}{c|c}
     1 + \gamma_f^{-1} \, \mathfrak{m} + \gamma_f^{-2} \mathfrak{p}_{+} \lr{1 + \gamma_f^{-1} \, \mathfrak{m}}^{-1} \mathfrak{p}_{-}  &
     \gamma_f^{-1} \mathfrak{p}_{+} \lr{1 + \gamma_f^{-1} \, \mathfrak{m}}^{-1} \\ \hline
     \gamma_f^{-1} \lr{1 + \gamma_f^{-1} \, \mathfrak{m}}^{-1} \mathfrak{p}_{-} &
     \lr{1 + \gamma_f^{-1} \, \mathfrak{m}}^{-1} 
    \end{array}
    } .
\end{eqnarray}
To discuss the relation of the operators $\alpha^{-1} \, \beta$ to fermionic transfer matrices of the form $T_{\tau} = e^{\dt \, h\lrs{\vec{A}\lr{\tau}}} \, e^{i \dt \, A_0\lr{\tau}}$, we note that each transfer matrix $T_{\tau}$ is a product of a Hermitian operator $e^{\dt \, h\lrs{\vec{A}\lr{\tau}}}$ and a unitary operator $e^{i \dt \, A_0\lr{\tau}}$. Operators $\alpha^{-1}_{\tau} \, \beta_{\tau}$ can also be brought into a form 
\begin{eqnarray}
\label{eq:alphainv_beta_radial_decomposition}
    \alpha^{-1} \, \beta = 
    e^{\dt \, \bar{h}^{WD} } \, \tilde{U}_{\tau, 0} ,
\end{eqnarray}
by using the matrix radial decomposition, whereby any matrix $A$ can be represented as $A = R U$, with $R$ being Hermitian and positive-definite matrix, and $U$ being unitary. The unitary factor $\tilde{U}_{\tau, 0}$ can be interpreted as an additional renormalization of the time-like holonomy factors $U_{\tau,0} \rightarrow \tilde{U}_{\tau, 0} U_{\tau, 0}$ in the time-ordered products of $\alpha_{\tau}^{-1} \, \beta_{\tau} \, U_{\tau,0}$ in (\ref{eq:det_final}). The effect of this factor is similar to introducing ``fat'' Polyakov loops \cite{Datta:1512.04892}. Likewise, the Hermitian and positive-definite factor $e^{\dt \, \bar{h}^{WD} } $ accounts for additional renormalization of the Wilson-Dirac Hamiltonian in the presence of time-like clover terms.

At very high temperatures, chromo-electric fields $\vec{\mathcal{E}}$ are Debye-screened. With high-temperature dimensional reduction and spectral reconstruction as the main motivation for this work, let us therefore consider a simplified case in which the chromo-electric fields $\vec{\mathcal{E}}$ are neglected and the operators $\mathfrak{p}_{+}$ and $\mathfrak{p}_{-}$ become equal to each other: $\mathfrak{p}_{+} = \mathfrak{p}_{-} = \mathfrak{p}$.

In this case, the representation (\ref{eq:alphainv_beta}) makes it clear that $\alpha^{-1} \, \beta$ is a Hermitian operator: diagonal blocks are Hermitian operators, and off-diagonal blocks are Hermitian conjugate to each other. Following \cite{Alexandru:1009.2197,Nakamura:1009.2149}, one can also demonstrate that eigenvalues of $\alpha^{-1} \beta$ come in pairs $\lr{\lambda, \lambda^{-1}}$. Given that the spectrum of $\mathfrak{p}$ and $\mathfrak{m}$ operators has a finite support, it is also clear from (\ref{eq:alphainv_beta}) that eigenvalues of $\alpha^{-1} \, \beta$ are positive for sufficiently small $\gamma_f^{-1}$. 

Expanding (\ref{eq:alphainv_beta}) in powers of $\gamma_f^{-1}$, it is also easy to show that 
\begin{eqnarray}
\label{eq:}
\alpha^{-1} \beta 
= 
1 + \gamma_f^{-1} h^{WD} + \gamma_f^{-2} 
\lr{
\begin{array}{cc}
 \mathfrak{p}^2 & -\mathfrak{p} \, \mathfrak{m} \\
 -\mathfrak{m} \, \mathfrak{p} & \mathfrak{m}^2 \\
\end{array}
}
+
O\lr{\gamma_f^{-3}} = e^{\gamma_f^{-1} \, h^{WD}} + O\lr{\gamma_f^{-2}} .
\end{eqnarray}
Given that $\gamma_f^{-1} \sim \dt/a_s$ and $e^{\dt \, h} = 1 + \dt \, h + O\lr{\dt^2}$, this establishes the connection between the factors $\alpha_{\tau}^{-1} \beta_{\tau}$ in (\ref{eq:det_final}) and the exponentials $e^{-\dt \, h}$ in (\ref{eq:3d_determinant_generic}), which are equivalent to $e^{+\dt \, h}$ by virtue of particle-hole symmetry of the Dirac Hamiltonian. 

Furthermore, since $\alpha^{-1} \beta$ is a Hermitian and positive-definite operator (for sufficiently small $\gamma_f^{-1}$), we can introduce an effective single-particle Hamiltonian $\bar{h}^{WD}$ that governs the Euclidean time evolution at finite temporal lattice spacing. To this end, we identify $\alpha^{-1} \beta = e^{\dt \, \bar{h}^{WD}}$. Taking the matrix log of both parts and expanding in powers of $\gamma_f^{-1}$, we obtain
\begin{eqnarray}
\label{eq:hWD_eff}
 \bar{h}^{WD} = h^{WD} + \gamma_f^{-1} \,
 \lr{
    \begin{array}{cc}
        \lr{\mathfrak{p}^2 - \mathfrak{m}^2}/2 & \lr{\mathfrak{p} \, \mathfrak{m} - \mathfrak{m}\, \mathfrak{p}}/2 \\
        -\lr{\mathfrak{p} \, \mathfrak{m} - \mathfrak{m}\, \mathfrak{p}}/2 & - \lr{\mathfrak{p}^2 - \mathfrak{m}^2}/2 \\
    \end{array}
 }
 + O\lr{\gamma_f^{-2}} .
\end{eqnarray}
We conclude that the effective Hamiltonian $\bar{h}^{WD}$ indeed converges to the Wilson-Dirac Hamiltonian $h^{WD}$ as defined in (\ref{eq:hWD_eff}) in the continuum time limit $\gamma_f^{-1} \rightarrow 0$.

To demonstrate that $T_{\tau}^{WD} = \alpha_{\tau}^{-1} \, \beta_{\tau} U_{\tau, 0}$ is indeed the transfer matrix that describes the propagation of physical fermion states, it is also useful to rewrite the fermion propagator $D_{WD}^{-1}$ in terms of three-dimensional transfer matrices. To this end, we write 
\begin{eqnarray}
\label{eq:DWD_inv_prelim}
    D_{WD}^{-1} = u_t \mathcal{P} \, \lr{u_t \, \gamma_0 \, D_{WD} \, \mathcal{P}}^{-1} \, \gamma_0 .
\end{eqnarray}
Due to its special ``upper-diagonal'' structure, the inverse of the operator $\lr{u_t \, \gamma_0 \, D_{WD} \, \mathcal{P}}^{-1}$ can be explicitly calculated as:
\begin{eqnarray}
\label{eq:g0_DWD_P_inv}
 \mathcal{G}_{\tau_1, \tau_2} = \lr{\lr{u_t \, \gamma_0 \, D_{WD} \, \mathcal{P}}^{-1}}_{\tau_1,\tau_2}
 =
 \begin{cases}
  -\mathcal{F}_{\tau_1} \prod\limits_{\tau = \tau_1}^{\tau_2 - 1} T_{\tau}^{WD} \, \alpha_{\tau_2}^{-1} , & \tau_2 > \tau_1 \\
  - \mathcal{F}_{\tau_1} \alpha_{\tau_1}^{-1}, & \tau_2 = \tau_1 \\
  \mathcal{F}_{\tau_1} 
  \prod\limits_{\tau=\tau_1}^{N_{\tau}-1} T_{\tau}^{WD} 
  \prod\limits_{\tau=0}^{\tau_2-1} T_{\tau}^{WD} \, \alpha_{\tau_2}^{-1}
  , & \tau_2 < \tau_1 \\
 \end{cases}  ,
 \nonumber \\
 \mathcal{F}_{\tau_1} = \lr{I + 
    \prod\limits_{\tau=\tau_1}^{N_{\tau}-1} T^{WD}_{\tau} 
    \prod\limits_{\tau=0}^{\tau_1-1} T^{WD}_{\tau}  }^{-1} .
\end{eqnarray}

Expressing $D_{WD}^{-1}$ in terms of $\mathcal{G} = \lr{u_t \, \gamma_0 \, D_{WD} \, \mathcal{P}}^{-1}$ as in (\ref{eq:DWD_inv_prelim}) and inserting the result into the general expression
\begin{eqnarray}
\label{eq:Gtau_general}
 G\lr{\tau}
 =
 - \vev{\tr\lr{\gamma_0 \, O \, \lr{D^{-1}_{WD}}_{0, \tau} \gamma_0 \, O \, \lr{D^{-1}_{WD}}_{\tau, 0}}}
\end{eqnarray}
for the connected part of the fermionic correlators, we obtain an intermediate expression
\begin{eqnarray}
\label{eq:GE_WD_general_prelim}
 G\lr{\tau}
 =
 - u_t^2
 \vev{\tr\lr{
    O \lr{\mathcal{P} \mathcal{G}}_{0,\tau}
    O \lr{\mathcal{P} \mathcal{G}}_{\tau,0}
 }} .
\end{eqnarray}
We now use explicit expressions (\ref{eq:g0_DWD_P_inv}) and (\ref{eq:P_aux}) for the block matrices $\mathcal{G}$ and $\mathcal{P}$ to obtain an expression of the form similar to (\ref{eq:bilinear_correlator_general}). We first represent $\lr{\mathcal{P} \mathcal{G}}_{0,\tau}$ as
\begin{eqnarray}
\label{eq:GE_WD_general_prelim1}
    \lr{\mathcal{P} \mathcal{G}}_{0,\tau}
    =
    P_{-} \mathcal{G}_{0, \tau}
    +
    P_{+} U_{0,0} \mathcal{G}_{1, \tau}
    =
    -P_{-} \mathcal{F}_{0} \prod\limits_{\tau' = 0}^{\tau - 1} T_{\tau'}^{WD} \, \alpha_{\tau}^{-1}
    -P_{+} U_{0,0}
    \mathcal{F}_{1} \prod\limits_{\tau' = 1}^{\tau - 1} T_{\tau'}^{WD} \, \alpha_{\tau}^{-1}
    = \nonumber \\  =
    - \lr{P_{-}  + P_{+} U_{0,0} \lr{T^{WD}_0}^{-1}} \mathcal{F}_{0} \prod\limits_{\tau' = 0}^{\tau - 1} T_{\tau'}^{WD} \, \alpha_{\tau}^{-1}
    \equiv 
    -\mathcal{R}_0 
    \mathcal{F}_{0} \prod\limits_{\tau' = 0}^{\tau - 1} T_{\tau'}^{WD} \, \alpha_{\tau}^{-1}
    , 
\end{eqnarray}
where we used the identity $\mathcal{F}_1 = \lr{T^{WD}_0}^{-1} \mathcal{F}_0 \, \, T^{WD}_0$ and the fact that $\mathcal{P}_{\tau, \tau'}$ is only non-zero for $\tau' = \tau$ and $\tau' = \tau + 1$, and introduced the notation
\begin{eqnarray}
\label{eq:Rtau_def}
  \mathcal{R}_{\tau} 
  = 
  P_{-}  + P_{+} U_{\tau,0} \lr{T^{WD}_\tau}^{-1} .
\end{eqnarray}
In this and the following derivation, we also assume that $1 < \tau < N_{\tau}-2$ for simplicity, to omit technically straightforward, but space-consuming considerations of boundary terms at small $\tau = 0, 1$.

We can transform $\lr{\mathcal{P} \mathcal{G}}_{\tau,0}$ in a similar way, using the general identity $\mathcal{F}_{\tau+1} = \lr{T^{WD}_{\tau}}^{-1} \mathcal{F}_{\tau} \, \, T^{WD}_{\tau}$:
\begin{eqnarray}
\label{eq:GE_WD_general_prelim2}
    \lr{\mathcal{P} \mathcal{G}}_{\tau,0}
    =
    P_{-} \mathcal{G}_{\tau, 0}
    +
    P_{+} U_{\tau,0} \mathcal{G}_{\tau+1, 0}
    = \nonumber \\ = 
    P_{-} \mathcal{F}_{\tau}
    \prod\limits_{\tau'=\tau}^{N_{\tau}-1} T^{WD}_{\tau'}
    \alpha_0^{-1}
    +
    P_{+} U_{\tau,0} \mathcal{F}_{\tau+1}
    \prod\limits_{\tau'=\tau+1}^{N_{\tau}-1} T^{WD}_{\tau'}
    \alpha_0^{-1}
    = \nonumber \\ = 
    \lr{P_{-}  + P_{+} U_{\tau,0} \lr{T^{WD}_\tau}^{-1}} \mathcal{F}_{\tau} \prod\limits_{\tau' = \tau}^{N_{\tau} - 1} T_{\tau'}^{WD} \, \alpha_{0}^{-1}
    \equiv 
    \mathcal{R}_{\tau} 
    \mathcal{F}_{\tau} \prod\limits_{\tau' = \tau}^{N_{\tau} - 1} T_{\tau'}^{WD} \, \alpha_{0}^{-1} .
\end{eqnarray}
Inserting the expressions (\ref{eq:GE_WD_general_prelim1}) and (\ref{eq:GE_WD_general_prelim2}) into (\ref{eq:GE_WD_general_prelim}), we obtain closed expressions of the form (\ref{eq:bilinear_correlator_general}), where $T_{\tau}^{WD}$ is used as the transfer matrix
\begin{eqnarray}
\label{eq:GE_WD_general}
 G\lr{\tau} = \vev{\tr\lr{\tilde{O}_0 \mathcal{F}_0 \prod\limits_{\tau'=0}^{\tau-1} T^{WD}_{\tau'} \tilde{O}_{\tau} \prod\limits_{\tau'=\tau}^{N_{\tau}-1} T_{\tau'}^{WD} \mathcal{F}_0 }} ,
\end{eqnarray}
where 
\begin{eqnarray}
\label{eq:operator_modification}
    \tilde{O}_{\tau} = u_t \, \alpha_{\tau}^{-1}
    O R_{\tau}
\end{eqnarray}
is a modified single-particle observable operator, which differs from the original single-particle operator $O$ in the fermionic bilinear operator $\hat{\mathcal{O}} = \psi^{\dag} O \psi$ by terms of order $\gamma_f^{-1}$ and higher.

In numerical studies of condensed matter systems, where relativistic invariance does not constrain the lattice discretization as much as in lattice QCD, the lattice artifacts due to imaginary time discretization can be to some extent absorbed into a modification of the Green-Kubo relations, where exponentials like $e^{-w \, \tau}$ are replaced by products of the form $\lr{1 - w \, \dt}^{\tau/\dt}$, see e.g. Eq.~(13) in the Supplementary Material of \cite{Ulybyshev:2104.09655}. The representation (\ref{eq:GE_WD_general}) demonstrates that for Wilson-Dirac fermions lattice artifacts cannot be absorbed into a simple redefinition of the Green-Kubo kernel or a transformation of the energy scales. The reason is the non-trivial mixing between forward and backward finite difference approximations for derivatives in the Euclidean time direction.

\section{Dimensional reduction at high temperatures}
\label{sec:dimensional_reduction_high_T}

After identifying $T^{WD} = e^{\dt \bar{h}^{WD}}$ and taking the high-temperature limit where the space-like link variables are independent of the Euclidean time $\tau$ and time-like links are equal to unity, we can map the expression (\ref{eq:GE_WD_general}) to the form (\ref{eq:bilinear_correlator_genera_high_temp}), with $h$ replaced by $\bar{h}^{WD}$. In this limit, real-time spectral functions 
\begin{eqnarray}
\label{eq:spectral_func}
 \bar{\rho}\lr{w} = \sum\limits_{m,n} \abs{\bra{\bar{m}} \tilde{O} \ket{\bar{n}}}^2 \frac{e^{-\beta \bar{\epsilon}_m} - e^{-\beta \bar{\epsilon}_n}}{\lr{1 + e^{-\beta \bar{\epsilon}_n}} \lr{1 + e^{-\beta \bar{\epsilon}_m}}}
 \frac{\delta\lr{w - \lr{\bar{\epsilon}_n - \bar{\epsilon}_m}}}{\bar{\epsilon}_n - \bar{\epsilon}_m}
\end{eqnarray}
that correspond to Euclidean correlators (\ref{eq:GE_WD_general}) calculated with Wilson-Dirac fermions will therefore contain eigenvalues and eigenvectors $\lrc{\bar{\epsilon}_n, \ket{\bar{n}}}$ of the modified single-particle Hamiltonian $\bar{h}^{WD}$, rather than the eigensystem of a ``bare'' Wilson-Dirac Hamiltonian $h^{WD}$ as defined in (\ref{eq:hWD}). In addition, this expression will contain matrix elements $\bra{\bar{m}} \tilde{O} \ket{\bar{n}}$ of the modified observable operator $\tilde{O}$, which becomes independent of $\tau$ in the high-temperature, static-field limit. 

\section{Conclusions}
\label{sec:conclusions}

The results derived in these Proceedings, in particular, the definition (\ref{eq:hWD_eff}) of the effective Hamiltonian at finite temporal time step, the explicit expressions (\ref{eq:GE_WD_general}) and (\ref{eq:spectral_func}) for fermionic Euclidean correlators and for the corresponding real-frequency spectral functions, could be used for direct analysis of lattice artifacts in the numerical spectral reconstruction procedure.

With the Green-Kubo relations 
\begin{eqnarray}
\label{eq:Green_Kubo}
 G_E\lr{\tau} = \int\limits_{0}^{+\infty} dw \, \frac{w \, \cosh\lr{w\lr{\tau - \beta/2}}}{\sinh\lr{w \beta/2}} \, \rho\lr{w}
\end{eqnarray}
being exponentially insensitive to low-frequency/late-time real-time spectral functions, very different spectral functions can result in very similar Euclidean correlators. Conversely, even small artifacts in Euclidean correlators may correspond to large artifacts in the reconstructed spectral functions. It is therefore important to understand potential effect of lattice artifacts on numerical reconstruction of $\rho\lr{w}$ in (\ref{eq:Green_Kubo}). Finite-volume artifacts in the context of spectral reconstruction have only recently been studied in \cite{Bresciani:2606.14349}.

To the best of our knowledge, artifacts due to time discretization are not well understood at the moment, and have never been systematically studied.  The framework developed in these Proceedings allows to explicitly quantify these artifacts. Explicit numerical analysis on realistic lattice QCD data will be the subject of a future work.

\acknowledgments{
 The work of P.~Buividovich was funded in part by the STFC Consolidated Grant ST/X000699/1.}


\begin{thebibliography}{10}
\expandafter\ifx\csname url\endcsname\relax
  \def\url#1{{\tt #1}}\fi
\expandafter\ifx\csname urlprefix\endcsname\relax\def\urlprefix{URL }\fi
\providecommand{\eprint}[2][]{\href{https://arxiv.org/abs/#2}{ArXiv:#2}}

\bibitem{Nakamura:1009.2149}
K.~Nagata, A.~Nakamura, {\em Wilson fermion determinant in lattice {QCD}\/}, Phys.~Rev.~D {\bf 82} (2010), 094027, \eprint{1009.2149}.
\href{http://dx.doi.org/10.1103/PhysRevD.82.094027}{\url{http://dx.doi.org/10.1103/PhysRevD.82.094027}}

\bibitem{Alexandru:1009.2197}
A.~Alexandru, U.~Wenger, {\em {QCD} at non-zero density and canonical partition functions with {Wilson} fermions\/}, Phys.~Rev.~D {\bf 83} (2011), 034502, \eprint{1009.2197}.
\href{http://dx.doi.org/10.1103/PhysRevD.83.034502}{\url{http://dx.doi.org/10.1103/PhysRevD.83.034502}}

\bibitem{Buividovich:25:3}
P.~V. Buividovich, B.~Hind, {\em Spectral reconstruction based on dimensional reduction in high-temperature gauge theories\/} (2025), \eprint{2512.23560}.

\bibitem{zubkov2018momentum}
M.~A. Zubkov, {\em Momentum space topology of {QCD}\/}, Annals of Physics {\bf 393} (2018), 264--287, \eprint{1610.08041}.

\bibitem{chernodub2017scale}
M.~N. Chernodub, M.~Zubkov, {\em Scale magnetic effect in quantum electrodynamics and the {Wigner-Weyl} formalism\/}, Physical Review D {\bf 96} (2017)~(5), 056006, \eprint{1703.06516}.

\bibitem{zhang2020influence}
C.~Zhang, M.~Zubkov, {\em Influence of interactions on the anomalous quantum {Hall} effect\/}, Journal of Physics A: Mathematical and Theoretical {\bf 53} (2020)~(19), 195002, \eprint{1902.06545}.

\bibitem{Kogut:75:1}
J.~Kogut, L.~Susskind, {\em Hamiltonian formulation of {W}ilson's lattice gauge theories\/}, Phys.~Rev.~D {\bf 11} (1975), 395.
\href{http://dx.doi.org/10.1103/PhysRevD.11.395}{\url{http://dx.doi.org/10.1103/PhysRevD.11.395}}

\bibitem{Blankenbecler:PhysRevD.24.2278}
R.~Blankenbecler, D.~J. Scalapino, R.~L. Sugar, {\em {Monte Carlo} calculations of coupled boson-fermion systems. {I}\/}, Phys.~Rev.~D {\bf 24} (1981), 2278.
\href{http://dx.doi.org/10.1103/PhysRevD.24.2278}{\url{http://dx.doi.org/10.1103/PhysRevD.24.2278}}

\bibitem{Buividovich:12:1}
P.~V. Buividovich, M.~I. Polikarpov, {\em {Monte-Carlo} study of the electron transport properties of monolayer graphene within the tight-binding model\/}, Phys.~Rev.~B {\bf 86} (2012), 245117, \eprint{1206.0619}.
\href{http://dx.doi.org/10.1103/PhysRevB.86.245117}{\url{http://dx.doi.org/10.1103/PhysRevB.86.245117}}

\bibitem{Wenger:2302.07385}
U.~Wenger, {\em {Transfer matrices and temporal factorization of the Wilson fermion determinant}\/}, PoS {\bf LATTICE2022} (2023), 042, \eprint{2302.07385}.

\bibitem{Futamura:1411.4262}
Y.~Futamura, S.~Hashimoto, A.~Imakura, K.~Nagata, T.~Sakurai, {\em {A filtering technique for the temporally reduced matrix of the Wilson fermion determinant}\/}, PoS {\bf LATTICE2014} (2014), 049, \eprint{1411.4262}.

\bibitem{Nagata:2014bra}
K.~Nagata, A.~Nakamura, S.~Hashimoto, {\em {A property of fermions at finite density by a reduction formula of fermion determinant}\/}, PoS {\bf LATTICE2013} (2014), 207.

\bibitem{Bilgici:2009gjc}
E.~Bilgici, J.~Danzer, C.~Gattringer, C.~B. Lang, L.~Liptak, {\em {Canonical fermion determinants in lattice QCD: Numerical evaluation and properties}\/}, Phys. Lett. B {\bf 697} (2011)~(1), 85--89, \eprint{0906.1088}.

\bibitem{Chen:hep-lat/0006019}
P.~Chen, {\em {Heavy quarks on anisotropic lattices: The Charmonium spectrum}\/}, Phys.~Rev.~D {\bf 64} (2001), 034509, \eprint{hep-lat/0006019}.
\href{https://dx.doi.org/10.1103/PhysRevD.64.034509}{\url{https://dx.doi.org/10.1103/PhysRevD.64.034509}}

\bibitem{Aarts:1412.6411}
G.~Aarts, C.~Allton, A.~Amato, P.~Giudice, S.~Hands, J.~Skullerud, {\em Electrical conductivity and charge diffusion in thermal {QCD} from the lattice\/}, JHEP {\bf 02} (2015), 186, \eprint{1412.6411}.
\href{http://dx.doi.org/10.1007/JHEP02(2015)186}{\url{http://dx.doi.org/10.1007/JHEP02(2015)186}}

\bibitem{OpenQCDfastsum}
J.~{Rylund Glesaaen}, B.~J\"{a}ger, {\em {openQCD-fastsum} code\/} (2018).
\href{https://gitlab.com/fastsum/openqcd-fastsum}{\url{https://gitlab.com/fastsum/openqcd-fastsum}}

\bibitem{Datta:1512.04892}
S.~Datta, S.~Gupta, A.~Lytle, {\em {Using Wilson flow to study the SU(3) deconfinement transition}\/}, Phys. Rev. D {\bf 94} (2016)~(9), 094502, \eprint{1512.04892}.

\bibitem{Ulybyshev:2104.09655}
M.~Ulybyshev, S.~Zafeiropoulos, C.~Winterowd, F.~Assaad, {\em {Bridging the gap between numerics and experiment in free standing graphene}\/} (2021), \eprint{2104.09655}.


\bibitem{Bresciani:2606.14349}
F.~A. Bresciani, M.~Bruno, M.~T. Hansen, {\em {Finite-volume effects on smeared spectral densities}\/} (2026), \eprint{2606.14349}.

\end{thebibliography}

\end{document}